\documentclass[a4paper]{article}

\usepackage{a4wide}
\usepackage{amsmath}
\usepackage{amssymb}
\usepackage{graphicx}

\def\adj{^\dagger}
\def\inv{^{-1}}

\def\bydef{ \stackrel{\mathrm{def}} =  }

\def\diag{\mathrm{diag}}
\def\cov{\mathrm{Cov}}

\def\E{\mathbb{E}}
\def\NN{\mathcal{N}}

\def\Pman{\mathbb{P}}
\def\Gman{\mathbb{G}}

\def\Sman{\mathbb{S}}

\newcommand\kld[2]{ {\textstyle D[\, #1 \|\,  #2 ]}}

\begin{document}

\title{Independent component analysis in the light of  Information Geometry}

\author{Jean-Fran\c cois Cardoso\\ Institut d'Astrophysique de Paris\\  CNRS, France}

\maketitle

  \begin{minipage}{1.0\linewidth}
    \large Version accepted for publication in `Information Geometry'.
    \\
    DOI: https://doi.org/10.1007/s41884-022-00073-x  
  \end{minipage}

\abstract{\sloppy I recall my first encounter with Professor Shun-ichi Amari who, once
  upon a time in Las Vegas, gave me a precious hint about connecting Independent Component
  Analysis (ICA) to Information Geometry.
  The paper sketches, rather informally, some of the insights gained in following this
  lead.}

\section{Amari and Pythagoras in Las Vegas}
\label{sec:APLV}

Independent Component Analysis (ICA) of a random $N$-vector $X$ consists in finding a
linear transform $B$ (an invertible $N\times N$ matrix) making the entries of $Y=BX$ `as
independent as possible'.
There are (infinitely) many matrices $B$ which can decorrelate the entries of $Y$ and, if
the data are Gaussian, decorrelation implies independence so that ICA has nothing to offer
here.
However, for \emph{non Gaussian} data, independence is stronger than decorrelation and the
situation is somehow the opposite: no matrix $B$ can produce a vector $Y=BX$ with
independent entries unless the distribution of $X$ is `special'.
That special case, of course, is when $X=AS$ where $S$ a vector of independent entries and
$A$ is some invertible matrix.
Moreover, in that case, there is an essential uniqueness of ICA: if the entries of $Y=BX$
are independent, they must be those of $S$, possibly up to permutation and rescaling or,
equivalently, $B$ must be of the form $B=PA\inv$ where matrix $P$ has one and only one
non-zero entry in each row and each column~\cite{Como94:SP}.

In other words, the only way of restoring independence of non Gaussian random variables
which have been mixed is to unmix them.
From this property stems the usefulness of ICA in many applications, whenever $N$ sensors
can be assumed to receive a mixture of independent sources but the coefficients of the
mixing are unknown or cannot be determined by physical modeling.
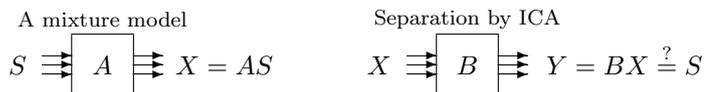
\begin{figure}[htb]
  \begin{center}
    \setlength{\unitlength}{0.4mm}
      \begin{picture}(280,32)(-40,-13)
        \put(30,15){\makebox(0,0)[c]{\footnotesize A mixture model}}
        \put(5,0){\makebox(0,0)[r]{$S$}}
        \put(10,3){\vector(1,0){10}}           
        \put(10,0){\vector(1,0){10}}
        \put(10,-3){\vector(1,0){10}} 
        \put(20,-10){\framebox(20,20){$A$}}
        \put(40,3){\vector(1,0){10}}           
        \put(40,0){\vector(1,0){10}}           
        \put(40,-3){\vector(1,0){10}}           
        \put(54,0){\makebox(0,0)[l]{$X=AS$}}
        \put(150,15){\makebox(0,0)[c]{\footnotesize Separation by ICA}}
        \put(125,0){\makebox(0,0)[r]{$X$}}
        \put(130,3){\vector(1,0){10}}           
        \put(130,0){\vector(1,0){10}} 
        \put(130,-3){\vector(1,0){10}}           
        \put(140,-10){\framebox(20,20){$B$}}
        \put(160,3){\vector(1,0){10}}          
        \put(160,0){\vector(1,0){10}}          
        \put(160,-3){\vector(1,0){10}}          
        \put(176,2){\makebox(0,0)[l]{$Y=BX  \stackrel{?}{=} S$}}
      \end{picture}
  \end{center}
  \caption{The hypothetical data model of linear mixture $X=AS$ and a separating matrix
    $B$ trying to recover the underlying sources in $S$.  If the entries of $S$ are
    statistically independent and non Gaussian, then matrix $B$ can restore the
    independence between the entries of $Y=BX$ \emph{only} by separating the sources, that
    is, the entries of $Y$ are those of $S$ (possibly up to rescaling and permutation).}
  \label{fig:directinverse}
\end{figure}
ICA makes `blind source separation' possible: this is the ability to recover mixed
underlying sources without resorting to any prior information on the system (matrix $A$),
resorting only to one key assumption: non Gaussian, statistically independent sources.
This ability is what made ICA such an attractive tool in many applications in which the
presence of independent sources is a strong but often plausible hypothesis.

\medskip

In 1995, I was not yet aware that the Earth was heading toward an environmental disaster
so I shamelessly flew to Las Vegas to attend a symposium on nonlinear theory (whatever
that means) where I presented a paper on the invariance of ICA.
I had been working extensively on ICA for a few years but, having stumbled upon Amari's
book~\cite{amari:LNS:1985}, I was trying to familiarize myself with Information Geometry
which I found to be a fascinating and inspiring vision.
At the end of my presentation, Amari stood up and made a kind comment.  I was
overjoyed: the Grand Master of Information Geometry was entering the field of ICA!

After the session, Amari invited me for a drink and generously shared an idea with me.
I already knew that the Kullback-Leibler divergence (KLD)
$\kld{P}{Q} = \int \log \frac{P(x)}{Q(x)} \ d P(x) $ from a distribution $P$ to another
distribution $Q$ gives rise to a Pythagorean theorem when used in conjunction with an 
finite-dimensional exponential family of distributions.
Amari pointed to me that the set of $N$-variate distributions with independent entries,
which is at the heart of ICA, can be seen as an exponential family, albeit an
infinite-dimensional one\footnote{We would need technical conditions to make this the more
  rigorous.  Hereafter, we assume that all source distributions have a strictly positive
  density with respect to the Lebesgue measure.}.

\medskip 

Indeed, consider the `product manifold' $\Pman$ as the set of $N$-variate probability
distributions which are the \textit{product} of their marginal distributions or, in other
words, distributions of $N$-vectors with independent entries.  Let $P_Y$ denote the
distribution of some random $N$-vector $Y$ and let $P_S=\prod_i P_{S_i}\in\Pman$ be any
distribution of independent entries.
By substitution, one easily finds 
\begin{equation}\label{eq:kullprod}
  \kld{P_{Y} }{ P_S}
  =
  \textstyle    \kld{P_Y}{ \prod_i P_{Y_i} }   +  \kld{\prod_i P_{Y_i}}{ \prod_i P_{S_i} }
\end{equation}
which shows that the minimum of $\kld{P_{Y} }{ P_S}$ over $\Pman$ is reached 
for $P_S= \prod_i P_{Y_i}$.
That minimum value is a well known quantity: the mutual information (between the entries)
of $Y$, denoted
\begin{equation} \label{eq:defI}
  \mathcal{I}(Y)
  \bydef
  \textstyle
  \kld{P_Y}{{\textstyle \prod_{i=1}^N  P_{Y_i} }} .
\end{equation}
Further, since
$\kld{\prod_i P_{Y_i}}{ \prod_i P_{S_i} }= \sum_i \kld{P_{Y_i}}{ P_{S_i} }$, the KLD from
an $N$-variate distribution $P_Y$ to a target distribution of independent components
$P_S=\prod_{i=1}^N P_{S_i}$ admits a decomposition:
\begin{equation}\label{eq:LLisIpM}
  \textstyle
  \kld{P_{Y} }{ \prod_i P_{S_i}} = \textstyle  \mathcal{I}(Y)   + \sum_i \kld{P_{Y_i}}{ P_{S_i} } 
\end{equation}
into a part $\mathcal{I}(Y)$ which does not depend on the target distribution $P_S$ but
only measures the amount of dependence between the entries of $Y$ and a part which only
measures marginal discrepancies between $Y$ and $S$.
Thus, Eq.~\ref{eq:LLisIpM} is the form taken by the Pythagorean theorem on $\Pman$.
With this insight, Amari provided me with a point of contact between ICA and Information
Geometry.

\medskip

The idea of using mutual information as a criterion for ICA had already been proposed in
the seminal paper of Comon~\cite{Como94:SP} but the geometrical connection could offer
more insights, in particular in relation to non Gaussianity.  Indeed, non Gaussianity is
not only required for blind separability but it can also be used as a criterion for
finding a separating matrix :
looking for maximally independent sources and looking for maximally non Gaussian sources
are two possible routes to blind separation~\cite{Como94:SP}.

This short paper shows how mutual information and non Gaussianity are geometrically
related.  Sec.~\ref{sec:LKM} follows the maximum likelihood principle for ICA, leading to
mutual information.  The latter is then related to non Gaussianity by another Pythagorean
theorem in Sec.~\ref{sec:depend-corr}, illustrating how non Gaussianity allows to express
statistical independence beyond mere decorrelation.  Some consequences for ICA are
sketched in Sec.~\ref{sec:fin}.

\section{Likelihood and Kullback matching for ICA}
\label{sec:LKM}
 
We start by setting up the simplest ICA model.  It assumes a zero-mean random $N$-vector
$S$ with independent entries ---the so-called `sources'--- mixed by an (invertible)
$N\times N$ matrix $A$ (the `mixing matrix'):
\begin{equation} \label{eq:ica}
  \textstyle
  X = AS  \quad\text{with}\quad S\sim Q(S)  = \prod_{i=1}^N q_i(S_i)
  \qquad
  \text{[the basic ICA model]}
\end{equation}
where $q_1, \ldots, q_N$ are $N$ scalar probability distributions for the sources.
The complete parameter set is $\theta=(A,Q)=(A , q_1, \ldots, q_N)$.  Since the aim of ICA
is to recover the sources by inverting $A$, the source distributions $q_i$ are considered
to be nuisance parameters while $A$ is the parameter of interest.

To gain some insights into the likelihood of the ICA model~(\ref{eq:ica}), we examine the
average shape of the log-density.
It is an easily demonstrated general fact that, for any parametric model
\begin{equation}\label{eq:MLisKM}
  \E_X \log P_\theta (X)  = - \kld{P_X}{P_\theta} - H(X) 
\end{equation}
where $H(X)$ denotes Shannon differential entropy.  Since the latter does not depend on
the model, the shape of the average log-likelihood landscape is controlled by
$\kld{P_X}{P_\theta}$, showing that the maximum likelihood principle corresponds to
minimizing the Kullback divergence from the data distribution $P_X$ to the model
distribution $P_\theta$.  
In the following, we explore the minimization of $\kld{P_X}{P_\theta}$ as the guiding
principle for ICA.

We can take advantage of a specific feature of the ICA model: it is a \emph{transformation
  model} and the KLD is invariant under invertible transforms.
Hence the KLD from the data distribution $P_X$ to the model distribution,$P_\theta$ of
$AS$ equals the KLD from the distribution of $A\inv X$ to the distribution of $S$ for any
invertible matrix $A$.
Therefore, for the ICA model~(\ref{eq:ica}), we have
\begin{equation}\label{eq:Smatch}
  \kld{P_X}{P_{\theta=(A, Q)}}
  = \kld{P_{A\inv X}}{P_{\theta=(I_N, Q)}}
  = \kld{P_{A\inv X} }{Q}
  .
\end{equation}
Since the data $X$ and the parameter of interest $A$ enter only via $Y=A\inv X$ in
Eq.~(\ref{eq:Smatch}), the message from the maximum likelihood principle is very clear:
the likeliest $A$ should make the transformed data $Y=A\inv X$ as close as possible to the
(hypothetical) source distribution $Q=\prod_i q_i$ in the sense of minimizing the Kullback
mismatch $\kld{P_{Y} }{Q}$.

Proceeding, we invoke decomposition (\ref{eq:LLisIpM}) which reads here as:
\begin{equation}
  \label{eq:both}
  \textstyle
  \kld{P_{A\inv X} }{Q}
  =
  \kld{P_Y}{\prod_i q_i}
  =
  \mathcal{I}(Y) + \sum_i \kld{P_{Y_i}}{q_i}
\end{equation}
and shows that minimizing $\kld{P_Y}{Q}$ is trying to achieve a composite objective:
making the entries of~$Y$ as independent as possible while also making their distributions
as close as possible to the marginal targets $q_1,\ldots, q_N$.

\medskip

Recall that the spirit of source separation is to proceed blindly as much as possible.
Just as we impose no constraints on $A$, it is desirable to let the nuisance parameters
$Q=\prod_i q_i$ be determined from the data themselves.
This is easily done (at least, in theory!) according to Eq.~(\ref{eq:both}): for any value
of $A$, the Kullback mismatch $ \kld{P_{A\inv X} }{\prod_i q_i}$ is minimized with respect
to the source distribution $q_i$ by making $\kld{P_{Y_i}}{q_i}$ equal to $0$,
\textit{i.e.} by estimating the source distribution $q_i$ to be the marginal distribution
$P_{Y_i}$.
Then we are left with
\begin{equation}  \label{eq:llToI}
  \min_{q_1,\ldots,q_N}
  \kld{P_{Y} }{\prod_i q_i}  = \mathcal{I}(Y) .
\end{equation}
We conclude that the maximum likelihood principle leads to the mutual information
$\mathcal{I}(Y)$ as the objective of choice for ICA when nothing is known about the source
distributions, in support of the original proposal of Comon~\cite{Como94:SP}.

\section{Independence and non Gaussianity}
\label{sec:depend-corr}

We already mentioned that looking for components which are
maximally non Gaussian is a possible route to source separation.  We now give the
geometric connection between these two objectives: moving away from being Gaussian and
moving closer to being independent.  All that is required are two applications of the
Pythagorean theorem.

\medskip

We start by defining a measure of non Gaussianity for a zero-mean%
\footnote{ For minimizing the notation and without any real loss of generality, all
  distributions are restricted to have zero mean in the following.} random $N$-vector with
distribution $P_Y$.
Denoting $\Gman$ the exponential family of all zero-mean $N$-variate Gaussian
distributions, the Pythagorean theorem on $\Gman$ takes the form
\begin{equation}\label{eq:pythG}
  \textstyle
  \kld{P_{Y} } {\NN(\Sigma) }  =  \kld{P_Y}{\NN(\cov Y)} +  \kld {\NN(\cov Y)}{\NN(\Sigma)} .
\end{equation}
where $\cov Y$ is the covariance matrix of $Y$ and where $\mathcal{N}(\Sigma)$ denotes the
zero-mean $N$-variate normal density with covariance matrix~$\Sigma$.
The non Gaussianity $\mathcal{G}(Y)$ of a zero-mean random vector $Y$ is naturally defined
as the divergence from its distribution to its best Gaussian approximation, which by
Eq.~(\ref{eq:pythG}), is $\NN(\cov Y)$:
\begin{displaymath}
  \mathcal{G}(Y) \bydef \kld{P_Y}{\mathcal{N}(\cov Y)} .
\end{displaymath}
Hence Eq. (\ref{eq:pythG}) shows that the divergence from a distribution to any Gaussian target has two parts: divergence
from Gaussianity (independent of the target) plus divergence of covariance matrices.

\medskip

Let us now combine the Pythagoras theorem of Eq.~(\ref{eq:kullprod}) related to
independence and the Pythagoras theorem of~(\ref{eq:pythG}) related to Gaussianity.
It is interesting to do it in terms of successive approximations.
When dealing with the distribution $P_Y$ of an $N$-vector which is too complicated to
handle, two widely used simplifying assumptions are that $Y$ is normally distributed or
that its entries are independent, that is, approximating distribution $P_Y$ either by
$P_Y^\Gman \bydef \mathcal{N}(\cov Y)$ or by $P_Y^\Pman \bydef\prod_i P_{Y_i}$.
In geometric terms, these approximations are projections onto $\Gman$ or onto $\Pman$.

An even cruder approximation would be to use both the Gaussian and the independent
approximations.
Projecting either $P_Y^\Pman$ onto $\Gman$ or projecting $P_Y^\Gman$ onto $\Pman$ leads in
both cases to $P_Y^{\Gman\Pman} \bydef \mathcal{N}(\diag(\cov Y))$.
\begin{figure}[htb]
  \centering
    \includegraphics{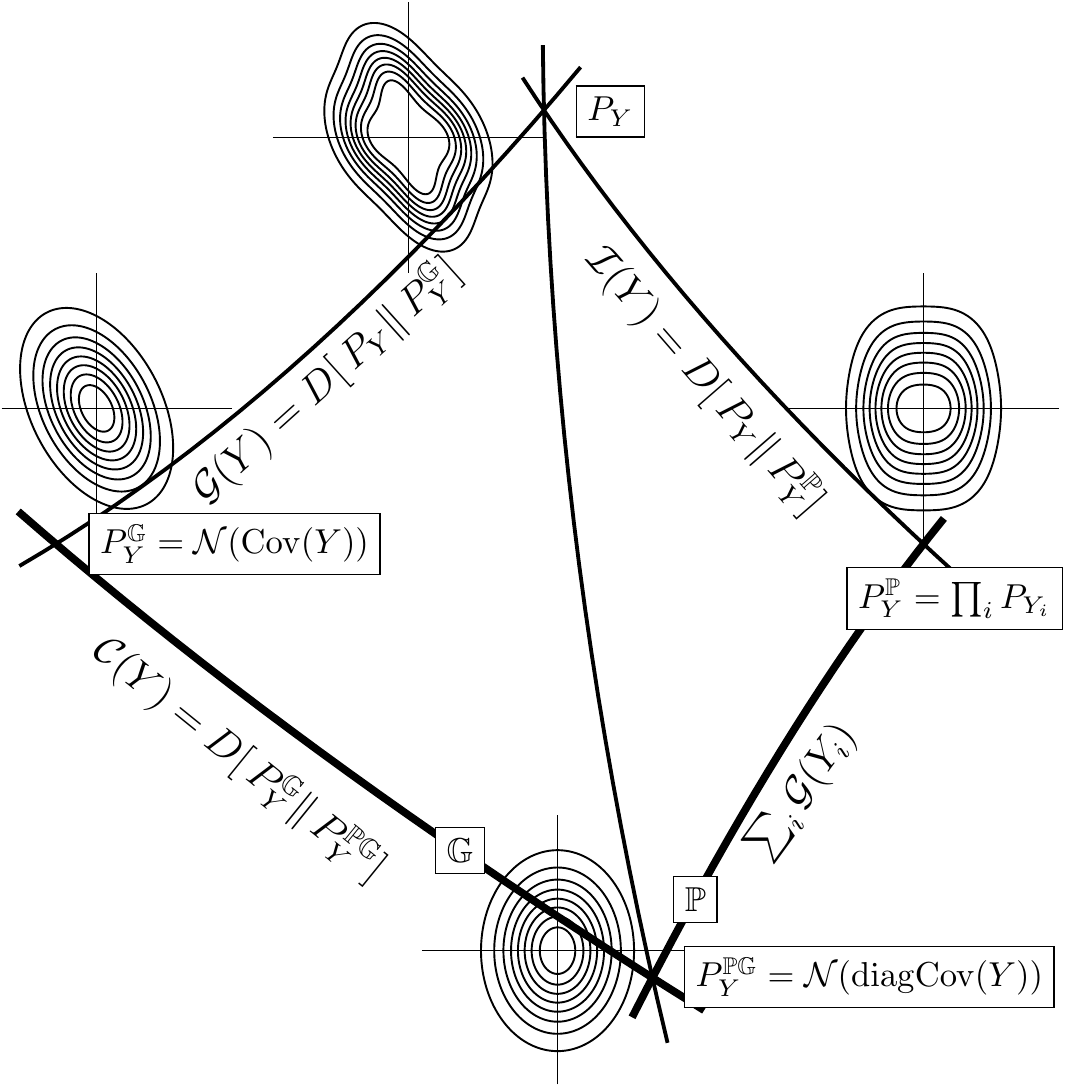}
  \caption{A probability density $P_Y$ for a vector $Y$ can be approximated as having
    independent components (approximation $P_Y^\Pman$) or as being Gaussian (approximation
    $P_Y^\Gman$) or both as $P_Y^{\Pman\Gman}$.  These approximations correspond to
    projections onto exponential manifolds.
    Those densities form two `right triangles', each giving rise to a Pythagorean theorem,
    and sharing a common hypotenuse, thus relating the `lengths' of the other sides,
    leading to Eq.~(\ref{eq:nice})
    The lengths of all sides have a clear and simple statistical meaning, allowing to
    connect independence, correlation and non Gaussianity in a single
    information-geometric picture.}
  \label{fig:DP}
\end{figure}
This is pictured in Fig.~\ref{fig:DP} showing the four aforementioned distributions,
forming two triangles: $[P_Y\rightarrow P_Y^\Gman\rightarrow P_Y^{\Pman\Gman} ]$ and
$[P_Y \rightarrow P_Y^\Pman \rightarrow P_Y^{\Pman\Gman}] $.
The key point is that these are two \emph{right} triangles which \emph{share a common
  hypotenuse} $[P_Y \rightarrow P_Y^{\Pman\Gman}] $.
Hence, $\kld{P_Y}{P_Y^{\Pman\Gman}}$ has two complementary expressions, using either
triangle:
\begin{equation}\label{eq:4sides}
  \textstyle
  \kld{P_Y}{P_Y^{\Pman\Gman}}  =
  \kld{P_Y}{P_Y^\Pman}  + \kld{P_Y^\Pman}{P_Y^{\Pman\Gman}}
  =
  \kld{P_Y}{P_Y^\Gman}  + \kld{P_Y^\Gman}{P_Y^{\Pman\Gman}}
\end{equation}
and applying each time the relevant Pythagorean relation~(\ref{eq:kullprod})
or~(\ref{eq:pythG}).

Two of the divergences appearing in (\ref{eq:4sides}) are already understood:
one is the mutual information $\mathcal{I}(Y)= \kld{P_Y}{P_Y^\Pman}$ measuring dependence;
the other is the non Gaussianity $\mathcal{G}(Y)= \kld{P_Y}{P_Y^\Gman}$ measuring\ldots\
just that.
The other two divergences also have a clear statistical meaning.
One is
$\kld{P_Y^\Gman}{P_Y^{\Pman\Gman}}=
\kld{\mathcal{N}(\cov Y)}{\mathcal{N}(\diag\cov Y)}$ measuring how far the covariance
matrix $\cov Y$ is from its diagonal part $\diag\cov Y$.
Hence, it measures the non diagonality of $\cov Y$ and therefore appears as the natural
scalar measure of the correlation between the entries of $Y$.
We thus define the \emph{correlation} of a random vector as
\begin{equation}
  \label{eq:defC}
  \mathcal{C}(Y) = \kld{P_Y^\Gman}{P_Y^{\Pman\Gman}}= \kld{\mathcal{N}(\cov Y)}{\mathcal{N}(\diag\cov Y)}
\end{equation}
The last divergence showing up in (\ref{eq:4sides}) is $\kld{P_Y^\Pman }{P_Y^{\Pman\Gman}}$.
Being a divergence between two distributions of vectors with independent entries, it is
just the sum of the pair-wise divergences between the entries.
Since each of those actually is the divergence from the distribution of $Y_i$ to its best
Gaussian approximation, one has
$\kld{P_Y^\Pman }{P_Y^{\Pman\Gman}}=\sum_i \mathcal{G}(Y_i)$, the sum of marginal
Gaussianities.
Thus Eq.~(\ref{eq:4sides}) finally yields the desired connection between mutual
information, correlation and non Gaussianity:
\begin{equation}\label{eq:nice}
  \mathcal{I}(Y)  + \sum_i \mathcal{G}(Y_i) = \mathcal{C}(Y)  + \mathcal{G}(Y) .
\end{equation}
The quantities defined via the KLD behave as nicely as possible:
by projection onto the Gaussian manifold $\Gman$, statistical dependence --- as measured
by mutual information $\mathcal{I}(Y) $--- reduces to correlation $\mathcal{C}(Y)$ while
by projection onto $\Pman$, the (full, joint) non-Gaussianity $\mathcal{G}(Y)$ is reduced
to marginal non-Gaussianity $\sum_i\mathcal{G}(Y_i)$.
Incidentally, the reduction of divergence is the same for both projections since
Eq.~(\ref{eq:nice}) also reads
$\mathcal{I}(Y) - \mathcal{C}(Y)= \mathcal{G}(Y) - \sum_i \mathcal{G}(Y_i)$.

\section{Relevance to independent component analysis}
\label{sec:fin}

The connection between independence, correlation and non Gaussianity of
Eq.~(\ref{eq:nice}) makes no reference to the ICA model and is independent of it.
Its impact on Independent Component Analysis is revealed by one final observation.
Recall that ICA deals with linear transforms of a vector $Y=A\inv X$.
Now, if a vector $Y$ undergoes some (invertible) linear transform, its Gaussian
approximation undergoes the \emph{same} transform.  Therefore, by invariance of the KLD,
the non Gaussianity $\mathcal{G}(Y)=\kld{P_Y}{P_Y^\Gman}$ is constant under linear
transforms.
Therefore, in a linear search for independent components, one has
\begin{equation}\label{eq:lucid}
  \mathcal{I}(Y)  = \mathcal{C}(Y) - \sum_i \mathcal{G}(Y_i)  + \text{constant} 
  \qquad
  \text{(for any $Y=BX$)}
  .
\end{equation}
Therefore, \emph{making the entries of $Y$ as independent as possible amounts to make them
  as uncorrelated and as non Gaussian as possible}, in the sense of Eq.~(\ref{eq:lucid})
\textit{i.e.} giving \emph{equal weight} to decorrelation and to non Gaussianity.

\medskip

The mutual information $\mathcal{I}(Y)$ is conceptually simple but quite a challenge to
estimate from data because density estimation is hard in the multidimensional case, and
downright impossible in practice as soon as the dimension $N$ is larger than a few units.
It is remarkable how relation (\ref{eq:lucid}) breaks down this complexity: the
correlation $\mathcal{C}(Y)$ is a simple function of a covariance matrix while each one of
the marginal non Gaussianities $\mathcal{G}(Y_i)$ only depends on the distribution of a
\emph{scalar} variable.  The only challenging term in (\ref{eq:lucid}) is hidden in the
constant and need not to be explicitly evaluated if one is only concerned with minimizing
the mutual information.

\medskip

This raises the algorithmic issue of actually minimizing the mutual information.
Since $\mathcal{I}(Y)$ itself, as a separation criterion, was obtained as a solution of
the minimization problem~\eqref{eq:llToI}, one approach is to alternate minimizations of
$\kld{P_Y}{\prod_i q_i}$ with respect to $A$ (changing $Y=A\inv X$) and with respect to
the source distributions $q_1,\ldots,q_N$.
It was shown in \cite{semipara} that the non-parametric estimation of the source densities
can be theoretically achieved without loss of statistical efficiency with respect to the
case when the source densities are known in advance.
This property has a geometric origin: the Fisher-orthogonality at point $Q=\prod_i q_i$
between the product manifold $\Pman$ and the $N^2$-dimensional `system manifold'
$\Sman\bydef\{ P_{CS} \, \vert \, C\in \mathrm{GL}(N), S\sim Q\}$
which is the manifold of all distributions of all invertible mixtures of $S$ when $P_S=Q$.

\medskip

In practice, a non-parametric estimation of the mutual information, or of the marginal non
Gaussianities or of the source densities could carry too much of a burden in many real
applications.  What happens if adopting the opposite option: choosing in advance some
model densities $q_i$ and keeping them fixed in the ICA likelihood?
Actually, the stationary points of $\kld{P_Y}{\prod_i q_i}$ with respect to linear
transforms of $Y$ have a simple expression: they are characterized by
$\E\{ \psi_i(Y_i) Y_j\} = 0$ ($1\leq i \neq j \leq N)$ where $\psi_i = -q_i'/q_i$ is the
\emph{score function} of density $q_i$.
These non linear decorrelation conditions are fulfilled if the entries of $Y$ are
independent because then $\E\{ \psi_i(Y_i) Y_j\} = \E\{ \psi_i(Y_i)\}\E\{ Y_j\} $ for
$i\neq j$ and the last factors $\E Y_j$ cancel for zero-mean sources.
Hence, independent sources in $Y$ are stationary points of $\kld{P_Y}{\prod_i q_i}$
\emph{regardless} of the choice of the source models $q_i$!

However, to find separated sources by minimizing $\kld{P_Y}{\prod_i q_i}$, one needs more
than a stationary point: one needs a minimum.
Whether or not $\kld{P_{A\inv X}}{\prod_i q_i}$ is at a local minimum with respect to
variations of $A$ depends on the guessed $q_i$ distributions being `not too wrong', a
condition which can receive a quantitative expression in terms of the correlation between
the true and guessed score functions $\psi_i$ \cite{amari97:_stabil,BssStabil}.
This robustness property could be traced back to a geometric property: the orthogonality
of $\Pman$ and~$\Sman$.

The robustness of ICA with respect to the source model is illustrated by the Infomax
algorithm \cite{BellS95} which is an important example since it triggered a lot of
interest for ICA in neurosciences.
Infomax uses a fixed, popular non-linear function $\psi_i(s)=\tanh(s)$ and tries to solve
$\E\{ \tanh(Y_i) Y_j\} = \delta_{ij}$.
Since $\tanh(s)$ is the score function for a density $q(s) \propto 1/\cosh(s)$ which has
much heavier tails than a Gaussian distribution, infomax will usually operate successfully
in uncovering sources with heavy-tailed distributions even if their density is not exactly
proportional to $1/\cosh(s)$ (albeit at the cost of some unavoidable loss of statistical
efficiency).
Using $\psi_i(s)=\tanh(s)$ is implicitly like trying to fit a model of heavy-tailed, or
\emph{sparse} sources.  We have seen that the best criterion $\mathcal{I}(Y)$ does not
specifically want sparse sources but rather non Gaussian sources and being sparse is just
a particular way of being non Gaussian.
%
In presence of sources with densities of various kinds, both heavy-tailed and
light-tailed, it becomes necessary to develop source-adaptive methods, in the spirit of
mutual information and of its decomposition~(\ref{eq:lucid}) in terms of decorrelation and
non-Gaussianity.

\medskip

A final comment is in order regarding the so-called `orthogonal' ICA methods.
This popular approach to ICA relies on the idea that source separation can proceed in two
steps: in a first easy step, the data are `whitened' (decorrelated and normalized to unit
variance) and in a second step they are rotated, hence preserving
decorrelation~\cite{apbook:2010}.
In other words, an orthogonal method seeks a separating matrix in the form
$B=U\, \cov(X)^{-1/2}$ where matrix $U$ is constrained to be a rotation ($UU\adj=I_N$).
Such a construction strictly enforces the decorrelation of $Y=BX$, \textit{i.e.}  it
guarantees $\mathcal{C}(Y)=0$.  Hence, it can be seen as a variant of mutual information
which would put an infinite weight on the objective of decorrelation, leaving only the
degrees of freedom in $U$ to express independence beyond decorrelation
by maximizing the marginal non Gaussianities $\sum_i G(Y_i)$.
Some loss of statistical efficiency is expected in the orthogonal approach since, as per
Eq.~(\ref{eq:lucid}), mutual information (which derives from the maximum likelihood
principle) wants to give equal weight to the objectives of decorrelation and of
`degaussianization'.

\section{Conclusion}

This paper focused on the geometrical connection illustrated by Fig.~\ref{fig:DP} and on
some of its consequences, so quite a few points were left unaddressed, in particular in
relation to ICA as a \emph{transformation} model.
That the parameter of interest $A$ lives in the multiplicative group $\mathrm{GL}(N)$ has
some nice consequences in terms of statistical and algorithmic performance.  In
particular, the natural gradient~\cite{amari:NatGrad} of Amari takes a very simple form in
ICA, where it becomes a `relative gradient'~\cite{CL-easi}.
But there is more geometry to ICA and the interested reader is referred to
\cite{geomindep} or \cite{lechapitre} for more on this topic.

\medskip

Information geometry offers a wonderful source of inspiration for scientists who like to
think in terms of pictures, graphs, sketches.  The connection between dependence,
correlation, and non Gaussianity presented in this paper can easily be demonstrated
without resorting to information geometry but I would never have uncovered it without
geometric thinking.  I am grateful to Professor Amari for starting it.

\section*{Declarations}

\par\noindent\textbf{Data Availability}: Data sharing is not applicable to this article as
no data sets were generated or analyzed during the current study.

\par\noindent\textbf{Conflict of interest}: The author states that there is no conflict of
interest.

\bibliographystyle{plain}
\bibliography{IGofICA}

\end{document}